\documentclass[twocolumn,aps,prl,showpacs,superscriptaddress]{revtex4-1}
\usepackage{lmodern}

\usepackage[latin9]{inputenc}
\usepackage{amsmath}
\usepackage{amssymb}
\usepackage{graphicx}
\usepackage{epstopdf}
\usepackage{color}
\usepackage{enumerate}
\usepackage{ulem}
\usepackage{soul}

\begin{document}
\bibliographystyle{apsrev}

\title{Non-equilibrium restoration of duality symmetry in the vicinity of the superconductor-to-insulator transition}

\author{I. Tamir}
\affiliation{Department of Condensed Matter Physics, The Weizmann Institute of Science, Rehovot 76100, Israel.}
\email{idan.tamir@weizmann.ac.il.; Corresponding author}
\author{A. Doron}
\affiliation{Department of Condensed Matter Physics, The Weizmann Institute of Science, Rehovot 76100, Israel.}
\author{T. Levinson}
\affiliation{Department of Condensed Matter Physics, The Weizmann Institute of Science, Rehovot 76100, Israel.}
\author{F. Gorniaczyk}
\affiliation{Department of Condensed Matter Physics, The Weizmann Institute of Science, Rehovot 76100, Israel.}
\author{G. C. Tewari}
\affiliation{Department of Condensed Matter Physics, The Weizmann Institute of Science, Rehovot 76100, Israel.}
\affiliation{Present Address: Department of Physics, Central University of Rajasthan,
Kishangarh, Ajmer 305 817, India.}
\author{D. Shahar}
\affiliation{Department of Condensed Matter Physics, The Weizmann Institute of Science, Rehovot 76100, Israel.}

\begin{abstract}
The magnetic field driven superconductor to insulator transition in thin films is theoretically understood in terms of the notion of vortex-charge duality symmetry. The manifestation of such symmetry is the exchange of roles of current and voltage between the superconductor and the insulator.
While experimental evidence obtained from amorphous Indium Oxide films supported such duality symmetry it is shown to be broken, counterintuitively,  at low temperatures where the insulating phase exhibits discontinuous current-voltage characteristics. Here, we demonstrate that it is possible to effectively restore duality symmetry by driving the system beyond the discontinuity into its high current, far from equilibrium, state.    
\end{abstract}

\maketitle

The superconductor to insulator transition \cite{goldmanpt51,physupekhi} (SIT) is an experimentally accessible quantum phase transition \cite{sondhirmp}. By varying an externally controlled parameter in the Hamiltonian, a disordered superconducting thin film can be driven between its superconducting and insulating ground states \cite{haviprl62, HebardPrl, Shaharprb, kapitulnikprl74, BaturinaJETP, goldmanprl94, bollingernat, bounatmat11}. Two decades ago Fisher theoretically studied \cite{fisherduality} a specific case in which an applied magnetic field ($B$) drives the SIT. At low $B$, the induced Abrikosov vortices are localized by the disorder and a superconducting state prevails. Upon increasing $B$, Fisher found that the proliferation of vortices can result in a Bose-Einstein condensation of the vortex state that, in turn, leads to insulating behavior where the Cooper-pairs are now localized \cite{kowal, BarberPRB, GantmakherJETP, craneprb751, YonatanNat, vallesprl103, FeigAnnals, KopnovPRL, ShermanPRL, sacepe2015}. The exchange of roles between the Cooper-pairs and vortices across the transition is analyzed via a duality transformation applied to the Hamiltonian \cite{FazioPrb}. 

Experimentally, vortex-charge duality will manifest itself via the exchange of roles of current ($I$) and voltage ($V$) between the superconductor and the insulator \cite{shahar1995, Mehta2012,dualityKap}. Duality symmetry implies that, for a given resistance ($R\equiv V/I$) measured at a given $B=B_{SC}$ in the superconductor, there exists a dual $B=B_{Ins}$ in the insulator where the conductivity ($G\equiv I/V$) obeys the condition $G(B_{Ins})=R(B_{SC})$.
In previous publications \cite{murthyEuro,MaozNat} we 
found that our data follow a phenomenological, power-law, form across the SIT:
\begin{equation}
R(B)=R_C(\frac{B}{B_C})^{P(T)}
\label{DualityEq}
\end{equation}
where $P(T)\sim \frac{1}{T}$, $B_C$ is the critical $B$ value of the SIT and $R_C\simeq R_Q=\frac{h}{4e^2}$ \cite{fisherprl90}. This functional form is duality symmetric:
The equality $G(B_{Ins})=R(B_{SC})$ holds whenever the condition $B_{SC}/B_C=B_C/B_{Ins}$ is fulfilled.

\begin{figure}
	\includegraphics  [width=8.6 cm] {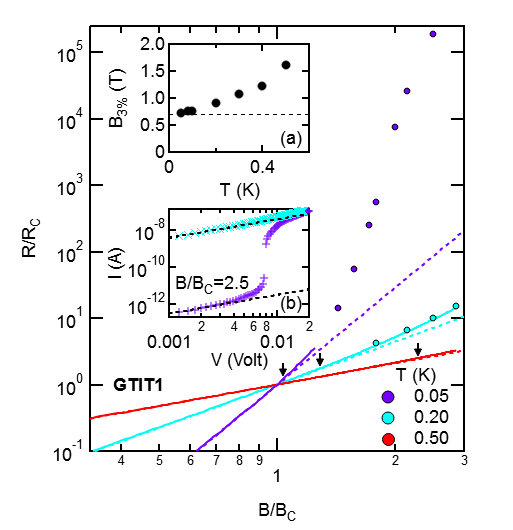}
	\caption{\textbf{(Color) Duality symmetry and its breakdown.} $R/R_C$ vs. $B/B_C$ (solid lines) measured at 0.05-0.5 K adopting a log-log graph. The dashed lines are power-law fits (in accordance with Eq. \ref{DualityEq}), in a range limited to the superconducting region, that are extended to the insulator. Deviations from the duality symmetry are indicated by arrows and are only observed in the insulating phase, resulting in a more resistive behavior. $R/R_C$'s extrapolated from full $I-V$'s are shown for $T=$0.05, 0.2 K data (circles). $R$ values are normalized with respect to $R_C=6.2\pm 0.6$ k$\Omega$. Note that deviations from duality symmetry at $T=$0.05 K exceeds three orders-of-magnitude. Inset (a): $B$ value of 3\% deviation from duality symmetry ($B_{3\%}$) vs $T$. The horizontal dashed line indicates $B_C=0.7$ T. (b): $I$ vs $V$ measured at $B=1.75$ T and $T=0.05, 0.2$ K ("+" and "X", respectively). The black dashed lines represent linear extrapolation to $V=0$.}  
	\label{RB}
\end{figure}

Counterintuitively, duality symmetry breaks down at low temperatures ($T$'s) \cite{MaozNat}. This is most conveniently illustrated through the deviations from the power-law dependence, graphically shown in Figure \ref{RB}. Interestingly, these deviations appear only in the insulating side of the SIT. In the superconducting side, the data continue to follow the power-law dependence down to our lowest $T$'s \cite{FeigelmanPhysC}. 

Together with the appearance of deviations from duality symmetry, our insulator develops strongly non-linear $I-V$ characteristics ($I-V$'s) \cite{sanprb53}. At $T\lesssim 0.2$ K, applying a bias $V$ above a well-defined $V=V_{th}$ (which is a function of both $B$ and $T$), results in a discontinuous increase, of several orders of magnitude, in $I$. 
Upon reducing $V$, a discontinuous decrease in $I$ is observed recovering previous $I$ values 
(see, for example, the 0.05 K data in Inset (b) of Figure \ref{RB} where a $I$ discontinuity is visible at $V\approx 7.5$ mV).

These discontinuities, initially associated with a new and exotic superinsulating phase \cite{vinokurnat}, were later theoretically linked \cite{borisprl} to a bi-stability of the electronic temperature ($T_{el}$).
Assuming: 1. Weak electron-phonon coupling, 2. Strong electron-electron interactions enabling self-thermalization to a well-defined $T_{el}$, 3. The Ohmic $R$ demonstrating insulating behavior, and 4. Linearity of the intrinsic $I-V$'s, whereby all deviations from linearity are associated with electron heating, Altshuler et al. \cite{borisprl} numerically solved the heat-balance equation $P\propto T_{el}^\beta-T^\beta$ and showed that $T_{el}$ can, at low enough $T$, either be near equilibrium, or at a significantly higher $T$ than that of the host lattice resulting in a far from thermodynamic equilibrium, high $I$ state. Several experimental results \cite{maozprl, doron2016, LevinsonPRB2016} support this approach. In what follows we will refer to the low $I$ ($V<V_{th}$), near equilibrium, regime as the high $R$ (HR) state, and to the high $I$ ($V>V_{th}$), out-of-equilibrium, regime as the low $R$ (LR) state.

We begin by following the low-$T$ evolution of the breakdown of duality symmetry in one of our amorphous Indium Oxide (a:InO) films, GTIT1 \cite{note}. In Figure \ref{RB} we plot $R/R_C$ vs. $B/B_C$, measured between 0.05-0.5 K, utilizing a 4-terminal Lock-In configuration (solid lines). We adopted a log-log graph to emphasize the power-law dependence of our data in accordance with Eq. \ref{DualityEq}. The dashed lines are extensions, to the insulating phase, of power-law fits done in a range limited to the superconducting phase. Deviations from the power-law dependence (indicated by arrows) and, consequently, from duality symmetry are observed only in the insulating phase, leading to higher $R$'s than those expected from duality symmetry. As $T$ is reduced, the starting point of these deviations approach $B_C$ (see Inset (a) of Figure \ref{RB}) and the deviations' magnitude increases. This trend becomes much more severe if we recall that standard 4-terminal measurements fail in the presence of non-linearities typical of insulators at low-$T$'s. This is apparent as we plot, alongside the 4-terminal data, $R$'s (normalized by $R_C$) evaluated from full $I-V$'s (circles). We note that, whenever the $I-V$'s are linear, the two measurement techniques are in agreement (0.2 K data in Figure \ref{RB} (b)). 

The severe breakdown of duality symmetry accompanies a transition to an insulating state that exhibits an unusual $T$-dependence. 
The significant upward deviations of our measured $R$'s reveal faster than activated behavior. This is supported by direct $R$ vs. $T$ ($R(T)$) measurements, in the insulating phase, near the SIT \cite{ovadia2015} where we showed that the $R(T)$'s not only exceeds activation, but seem to approach $R=\infty$ at a finite $T$.

\begin{figure}
	\includegraphics  [width=8.6 cm] {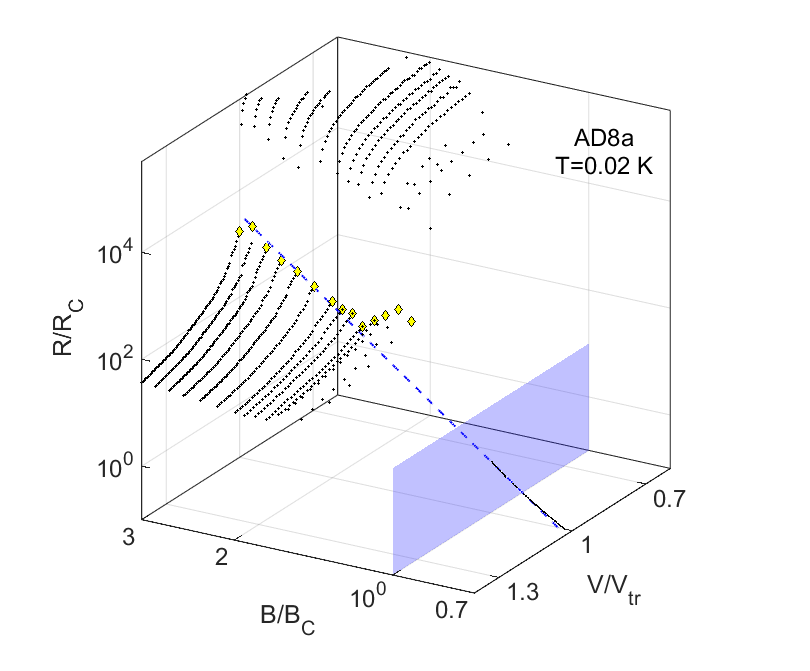}
	\caption{\textbf{Non-equilibrium restoration of duality symmetry.} Log-log plot of $R/R_C$ vs $B/B_C$ measured at $T=0.02$ K in the superconducting phase (solid line, right to the blue plane) and a power-law fit extended into the insulating phase (dashed line, left to the blue plane). Both $R$ and the fit, which where measured at zero bias $V$, are shifted to $V/V_{th}=1$. Superimposed are data of $\overline{R}/R_C$ vs $V$ measured at constant $B$'s and $T=0.02$ K (black circles). The yellow diamonds mark $\overline{R}(V_{th}^+)/R_C$.} 
	\label{duality}
\end{figure}

We now show how duality symmetry is restored by driving the system into the LR, out-of-equilibrium, state. This is demonstrated in Figure \ref{duality} where we plot $R/R_C$ vs. $B/B_C$ measured in the superconducting phase (solid line), extended to the insulating phase via fitting to a power-law (dashed blue line). Both $R$, which was measured at zero bias $V$, and the fit are shifted to $V/V_{th}=1$. In the insulating phase we superimpose the discontinuous $V/I$ ($\equiv \overline{R}$ \cite{note2},  normalized by $R_C$) data measured at constant $B$'s while decreasing $V$ (black circles). Both $R$ and $\overline{R}$ are measured at $T=0.02$ K \cite{note1}. 
In the LR ($V>V_{th}$), out-of-equilibrium, state, $\overline{R}$ gradually increases as we decrease $V$ up to a maximum value measured at $V=V_{th}^+$. The values $\overline{R}(V_{th}^+)$ (measured in the LR state), indicated by yellow diamonds, coincides with the extended power-law fit of the superconducting data up to 3 times $B_C$, restoring duality symmetry. Due to the discontinuous nature of the $I-V$'s, any further reduction of the applied $V$
 will result in a transition to the HR state and an orders of magnitude increase in $\overline{R}$.
We note, that at relatively low $B$'s, $\overline{R}(V_{th}^+)$ is not consistent with $R$ expected from duality symmetry. We attribute such deviations to the sensitivity of the HR branch near $B_C$ discussed in Ref \cite{AdamCat}.  

The restoration of duality symmetry can not be accommodated within currently available theoretical models. Duality symmetry, inherent to the superinsulating model \cite{vinokurnat}, is predicted for Ohmic transport at $V<V_{th}$, while we observe the opposite. Adopting the over-heated electrons framework \cite{borisprl} also leads to an apparent contradiction. As we have shown above, duality symmetry is most clearly evident in isotherms of $R(B)$, which follow a power-law (Figure \ref{RB}). Since the power-law is effectively restored immediately following the jump it seems reasonable that, at $V=V_{th}^+$, $T_{el}(V_{th}^{+})=T$. This is contradictory to the over-heated electrons framework where we expect that driving the system beyond $V_{th}$ into the LR state induces a significant increase of $T_{el}$ with respect to $T$ \cite{borisprl}, resulting in $T_{el}(V_{th}^{+})>>T$. If we adopt a view where data following the power-law $R(B)$ is necessarily isothermal, we have to conclude that
$R(V<V_{th}^{-})$, which is much greater than $R(V>V_{th}^{+})$, is more reasonably seen as being at $T_{el}<T$, as if the electronic system had condensed into a reduced-entropy state. This possibility calls for more experimental and theoretical studies.

\begin{figure}
	\includegraphics  [width=8.6 cm] {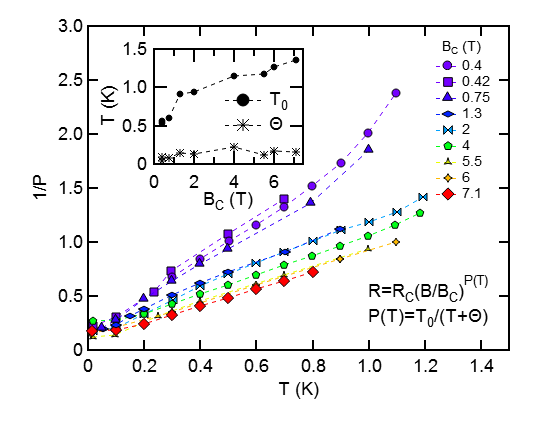}	
	\caption{\textbf{Power Law $T$ dependence.} $P^{-1}$ vs $T$ for several samples varying in size and disorder. $P$ is extracted by fitting $R(B)$'s measured at different $T$'s, in a range limited to the superconducting region, with a power law. The color scale represent $B_C$ of each sample from 7.1 T in red to 0.4 T in purple. Inset: $T_0$ and $\theta$ vs $B_C$ calculated by a linear fit of the data in the main figure (see Eq. \ref{PT}).}
	\label{PLT}
\end{figure}

Aside from duality symmetry, another analogy can be made between the LR, out-of-equilibrium, state and the superconducting phase: at low $T$'s, $R$ becomes weakly $T$-dependent. In the LR state, this $T$-dependence was previously discussed in Ref. \cite{borisprl}. In the superconducting phase, the experimental data deviates from the expected behavior of $P(T)\propto 1/T$ \cite{murthyEuro} (introduced in Eq. \ref{DualityEq}), and exhibits a more elaborate $T$-dependence. This observation is in compliance with recent reports of a possible metallic state intervening between the superconducting and insulating phases of disordered superconductors 
\cite{Ephronprl,MasonPRL,PRBYong,ClaireJoP,tsen2015nature,couedo2016}.
While at high $T$'s $P(T)\sim \frac{1}{T}$, fitting the full range of our data reveals a different dependence
\begin{equation}
P(T)=\frac{T_0}{T+\theta}
\label{PT}
\end{equation}
where both $\theta$ and $T_0$ are sample dependent parameters. In Figure \ref{PLT} we plot $P^{-1}$ vs $T$, for several different a:InO samples varying in size and disorder, visualizing the non-zero crossing of $P^{-1}$ at $T=0$. The color scale represent $B_C$ of each sample from 7.1 T in red to 0.4 T in purple. $P$ is extracted by fitting $R(B)$ data measured at the superconducting phase with a power-law. From the data we obtained $T_0$ and $\theta$ which are plotted in the inset of the figure as a function of $B_C$. Consequently, one observes a weak $R(T)$ dependence whenever $T\lesssim \theta$. We are not aware of any theoretical prediction of such functional behavior. Interestingly, a similar phenomenon was reported in the past near the quantum Hall-to-insulator transitions \cite{shahar1998}. A broader discussion regarding this behavior will follow in an upcoming publication.

In summary, we showed that the duality symmetry, observed at $T\geq 0.5$ K, does not describe the low $T$ physics of the $B$ driven insulating phase bordering the SIT and that experimental evidence point to the existence of a unique low $T$'s insulating phase. The physical nature behind this state is not yet understood and awaits further research. We also showed that this state is fragile and that duality symmetry, which is related to the continuation of the superconducting phase into the bordering insulating phase, can be restored at low $T$'s by driving the system out of equilibrium. The restoration of duality may point to an intriguing interplay between the insulating behavior at high $T$'s and the LR, out-of-equilibrium, state measured at low $T$'s. 

\section*{Acknowledgments}
We are grateful to B. Altshuler, M. Feigelman, V. Kravtsov and K. Michaeli for fruitful discussions. This research was supported by The Israel Science Foundation (ISF Grant no. 751/13) and The United States-Israel Binational Science Foundation (BSF Grant no. 2012210).

\bibliography{s1ahir}

\begin{thebibliography}{48}
\bibitem[{not({\natexlab{a}})}]{note}
\bibinfo{note}{GTIT1 is a 30nm, a:InO, hall bar shaped film undergoing the $B$
  driven SIT at $B_C=0.7$ T.}
\bibitem[{not({\natexlab{b}})}]{note2}
\bibinfo{note}{Within the over-heated electrons framework \cite{borisprl}, it
  is assumed that $R=R(T_{el})=\overline{R}$ for any $V$.}
\bibitem[{not({\natexlab{c}})}]{note1}
\bibinfo{note}{The data in Figure \ref{duality} were obtained from sample AD8a
  which is a 30nm, a:InO, hall bar shaped film undergoing the $B$ driven SIT at
  $B_C=0.6$ T.}
\end{thebibliography}

\end{document}